\documentstyle[aps,prb,epsf,multicol]{revtex}
\begin{document}
\draft
\title{Connection between 
   charge transfer and alloying core-level shifts
   based on density-functional calculations}

\author{M. Methfessel}
\address{
 Institute for Semiconductor Physics (IHP), Walter-Korsing-Str. 2, 
D-15230 Frankfurt(Oder), Germany}
\author{Vincenzo Fiorentini and Sabrina Oppo}
\address{Istituto Nazionale per la Fisica della Materia --
Dipartimento di Fisica, Universit\`a di  Cagliari, Italy}
\date{to appear on Phys. Rev. B, 15-February-2000}

\maketitle
\begin{abstract}
The measurement of alloying core-level binding energy (CLBE)
shifts has been used to give a precise meaning to 
the fundamental concept of charge transfer.
Here, {\it ab-initio} density-functional calculations 
for the intermetallic compound MgAu are used to
investigate models which try to
make a connection between the core levels shifts and charge transfer.
The calculated CLBE shifts agree well with experiment, and 
permit an unambiguous separation into initial-state and
screening contributions. Interestingly, the screening
contribution is large and cannot be neglected in any reasonable
description. Comparison of the calculated results with
the predictions of simple models show that these models are not adequate
to describe the realistic situation. On the positive side,
the accuracy of the density-functional calculations 
indicates that the combination of experiments
with such calculations is a powerful tool to investigate 
unknown systems.
\end{abstract}

\begin{multicols}{2}

\section{Introduction}   
The concept of charge transfer is fundamental to chemistry and 
condensed-matter physics. Unfortunately, it is frustratingly difficult to
give a precise definition of charge transfer, or even a well-defined
prescription for measuring it. This is equally true for related
quantities such as electronegativity and bond ionicity.

There have been a number of attempts to relate the charge transfer 
in an alloy to the positions of the core levels. These energies can
be measured with high accuracy using X-ray spectroscopy.\cite{egel}
In general terms, the core levels of an atom are shifted
when the atom's environment changes.
Interesting cases are, for example, when a crystal is formed out of 
free atoms,\cite{wil} when an atom is at the
surface\cite{zangwill,joha,scls,alden} rather than in the bulk of a
solid, or when an alloy is formed out of two elemental
solids.\cite{wert,darrah} 

In the case of the alloy core-level shift (the subject of this paper)
a major objective has been to find a well-defined connection
between the measured shift and the charge transfer between 
the constituents. Clearly, when charge
is moved from one atom to another, an electrostatic potential 
builds up, which modifies the energy needed to eject an electron 
from a core level into the vacuum. 
In the simplest form of the ``potential model'', 
the change of the potential felt by a core electron
is described using a Madelung term and an on-site contribution.\cite{darrah}
Unfortunately, cancellation between these effects and uncertainty in the 
the model parameters make it difficult to extract reliable 
charge transfers using this approach.

Furthermore, the simple potential model is valid only for 
the ``initial state'' picture, {\it i.e.} when describing 
the positions of the core levels in the alloy and the pure metal
{\it before} a core electron is removed. To compare with
the measured binding energies, a final-state screening contribution
must be taken into account: after a core hole is created, the 
remaining electrons relax to screen the hole. The kinetic energy
of the emitted electron includes the screening energy.
This can be included in the formalism
(by a term generally denoted $\delta R$), but this adds 
yet another parameter whose numerical value is poorly known.

An alternative procedure is described in Ref. \onlinecite{darrah}.
It was pointed out that the relaxation
energy can be measured directly via the shift of the Auger parameter 
(the sum of core-level ionization and Auger-energy shifts).
For the compounds AuMg and AuZn, the resulting values of $\delta R$
are inconsistent with the estimates used in an earlier work\cite{wert} 
based on the potential model, even though plausible ionicities were 
deduced there. A modified potential model was presented,\cite{darrah}
which relates the valence charge transfer in a metal to an atomic
property, namely the change in the potential at the core 
due to changes in the valence and core occupation numbers.
 
Under these circumstances, it is desirable to make a detailed
theoretical analysis of a typical system using a method which can quantify
the various contributions unambiguously. Here,\cite{prev}
we use {\it ab-initio} density-functional total-energy
calculations to study the MgAu alloy. Using a supercell technique,
the Mg, Au, and MgAu metals with and without a core hole on
a selected atom can be described accurately. This cleanly separates the 
core-level shift into initial-state and final-state
relaxation contributions, which can then be checked against the
appropriate models. In addition, direct inspection of the 
densities of states, on-site charges, and screening charge distributions
gives an understanding of the effects of alloying and the
different screening responses to a core hole.

>From the results, we are led to the conclusion that the final-state
relaxation contribution is not small, as
 it changes both the sign and the magnitude of the 
core-level shift on the Mg atom. This happens because the screening
of the Mg 1$s$ core hole is substantially less effective in the
alloy than in the pure Mg metal. In contrast, the relaxation
energy is found to be almost identical in the pure metal and the
alloy for Au. By inspecting the screening
density and comparing to free atom calculations, we try
to offer a simple explanation for the changes in the
screening properties upon alloying.

After obtaining a realistic picture of both the initial-state and
the final-state screening terms, we have tried to interpret these
results in terms of the potential model\cite{darrah}. 
However, is has not been possible to reproduce the features of the model.
In brief, the potential model 
assumes the following connection between the shift $\Delta V$ in the 
core-level binding energy and the charge transfer $\Delta q$:
\begin{equation}
  \Delta V =  ( k - M) \Delta q   \label{pot1}
\end{equation}
where $k\Delta q$ is the on-site Coulombic potential 
and $-M\Delta q$ is a Madelung term from charges on the other lattice sites. 
Sometimes additional terms are included,
{\it e.g.} in the form\cite{wert,friedman}
\begin{equation}
 \Delta V = k\Delta q - M\Delta q -\delta R -e \delta \phi~. \label{pot2}
\end{equation}
where $-\delta R$ is the change in the 
final-state relaxation energy and $-\delta\phi$ is the 
change in the Fermi energy. Our calculations can supply 
reliable values for the well-defined quantity $\delta R$, 
but cannot assign values to poorly-defined
quantities such as $\delta \phi$, $k$, and $M$.
A substantial effort to recover the potential model in either
formulation did not lead to any quantitative or even qualitative
agreement.

\section{Calculation and interpretation of core-level shifts} 
\label{back1}

Certain excitation energies (such as core-level shifts
and atomic ionization energies)
can be obtained as the difference of the total energies of two self-consistent 
density-functional calculations for the ground state. This can be done
whenever the excited state 
is formally the ground state for a different set of quantum 
numbers\cite{gj}. Although the 
calculated eigenvalues should not be
directly associated with the excitation energies, 
a connection can be made using Slater's transition state concept.\cite{slater}

During an experiment such as XPS,
an electron is emitted from the core state into the vacuum.  
The core-level binding energy
is the difference of the total energies between the unperturbed,
homogeneous crystal and the impurity system in which a single atom 
has a reduced core occupation. The first system is easy to handle
using standard band-structure techniques, whereas the second requires
some treatment suitable for impurities, such as use of supercells.

In a metal, a valence electron moves in 
from the surrounding crystal to screen the positive charge of 
the core hole. Effectively, the core electron has been lifted 
to the Fermi level hereby. The energy needed
to do this can be expected to depend on the position of the core 
eigenvalue before the excitation and on the degree of screening 
of the core hole.\cite{mahan} The separation 
into ``initial state'' and ``final-state screening'' 
contributions can be made clearer
using the transition-state concept. Within DFT, 
this is done using Janak's formula,\cite{janak} 
which states that the derivative of the total energy 
respective to a some occupation number equals the corresponding eigenvalue. 
Applied to the present situation, the charge $x$ is taken from the core state 
of one atom in the supercell and put into 
the valence band, and
\begin{equation}
 \frac{\partial E_T(x)}{\partial x} = E_{\rm F} - E_c (x) \equiv
\epsilon_c(x), \label{jak}
\end{equation}
where $E_T$ is the total energy, $E_{\rm F}$ is the Fermi energy, 
and $E_c$ is the core-level eigenvalue.

Actual calculations show that, to a very good approximation, the
core eigenvalue drops in a nearly linear fashion as it is deoccupied,
even though the overall core-level drop is substantial.
For example, $\epsilon_c$ increases from 1248.6 to 1364.7 eV
when the Mg $1s$ occupation is reduced from two to one.
Similarly, the Au $4f$ state starts at 78.6 eV below the Fermi energy
and drops to 93.3 eV. 

Assuming a strictly linear dependence of $\epsilon_c$ on $x$, 
the corel-level binding energy (the
change in total energy when one electron is taken from the core)
can be written in various illuminating ways:
\begin{eqnarray}
  E_T(1)-E_T(0) &=& \int_0^1 \epsilon_c(x) dx  \label{ts1}\\
        &\approx& \epsilon_c({\textstyle\frac{1}{2}}) \label{ts2}\\
        &\approx& {\textstyle\frac{1}{2}}
                [\epsilon_c(0)+\epsilon_c(1)]\label{ts3}\\
        &\approx& \epsilon_c(0) + {\textstyle\frac{1}{2}}[\epsilon_c(1)
                -\epsilon_c(0)]\label{ts4}
\end{eqnarray}
These equations express the full core-level binding energy (CLBE)
including final-state relaxation effects
in terms of the eigenvalues at different occupations. 
Eq.~\ref{ts2} is Slater's transition-state rule, and Eq. \ref{ts3}
shows that the CLBE is the average of the eigenvalues before
and after removing the core electron.  In Eq. \ref{ts4}, the first 
term $\epsilon_c(0)$ is the initial-state CLBE and the
relaxation contribution is identified as one-half of the core-eigenvalue
drop upon depopulation. 

This description is a useful tool to interpret the calculated results
because it makes contact between 
core-level shifts and differences in the screening response.
The core-level shift is the difference of the CLBE in two 
different environments, say $A$ and $B$. The initial-state
core-level shift is the difference of the static core levels
in the unperturbed systems. According to Eq. \ref{ts4}, the final-state 
relaxation contribution can be expressed as the difference of the 
core-eigenvalue drop upon depopulation. 
In general terms, a core level drops more strongly when
the valence electrons screen the core hole less efficiently.
Thus, the initial-state picture for the core-level shift is applicable
if the screening of the core hole is the same in both systems.
If there is a positive relaxation contribution 
to the core-level shift from $A$ to $B$, this shows that the
core-level drop is larger and the screening less effective
in system $B$. Conversely, a negative relaxation contribution
indicates that screening is more effective in $B$.

\section{Calculational procedure}  
To determine the initial-state CLBE, a calculation for the
unperturbed periodic systems is adequate. Hereby it is 
advantageous to use an all-electron method, which gives 
the core eigenvalues directly. For the complete CLBE including
final-state relaxation, a supercell is used and the difference
of the total energy with and without a core hole 
on one atom is evaluated. As discussed, the electron taken from
the core state is placed into the valence band. 
For additional information, the promoted charge can 
take non-integer values. Within a self-consistent DFT calculation, 
the important screening effects should be described accurately. 
Note that the properties of the surface,
specifically the work function, do not enter either description. This
must be the case for an acceptable model since the true core binding
energy relative to the Fermi level, expressed as the difference of 
two total energies, is a bulk property.

The electronic-structure and total-energy calculations 
presented here were done 
with the all-electron full-potential LMTO method,\cite{fp} within the
local approximation (LDA) to density-functional theory.\cite{gj}
Minimization of the energy under a constrained core occupation
is rigorously justified\cite{dede} in the DFT framework: 
a self-consistent calculation under the chosen constraint
provides a variational total energy in the parameter subspace
identified by the constraint.

We applied this technique to the core levels of the Mg
1$s$ and Au 4$f$ levels, first for the pure materials and
second for the binary MgAu alloy in the CsCl structure.
Accurate experimental data\cite{wert,darrah} exist on the core level 
shifts upon formation of this alloy.
To make comparisons between the materials
easier, the fcc structure was adopted for pure Mg. 
To study the  core-hole--excited solids, we used 
 16-atom supercells for both CsCl-structure
MgAu, and fcc Mg and Au. The distances of the core hole from its
periodic images exceeds 12 bohr in all cases, and   tests show that our
values for the core level shifts are converged with respect to
cell dimension. The localization of the calculated
density response to the core-hole perturbation, as detailed below,
provides an {\it a posteriori} justification for the used supercells.
The Brillouin-zone integration was done using more than 50
irreducible special points.
Muffin-tin radii for Mg respectively Au are 2.94 and 2.60 bohr
in the pure metals and 2.50 and 2.70 in the compound at the 
experimental lattice constants, and are scaled with the
lattice constant. All the calculations are scalar-relativistic and use
the Vosko {\it et al.} parameterization of the LDA 
exchange-correlation potential.\cite{vosko} 

\section{Results for  M\lowercase{g}A\lowercase{u}} 
\label{mgau-sec}

The calculated structural parameters for MgAu, Au and Mg are given 
in Table \ref{t1}. The results for these bulk systems are of standard
DFT-LDA quality. For the calculation of the core holes, supercells
were built up at the theoretical lattice constant.

Before moving to a discussion of the core level shifts, we point out
that the absolute core binding energies in Table \ref{t3}
(referred to $\rm E_F$, and 
obtained as total-energy differences) are in remarkable agreement 
with experiment, showing errors below 1\%.
The data in the Table
 also illustrates the large drop of the core eigenvalues when
an electron is removed (about 14.5 eV for Au and 120 eV for Mg).
By averaging the eigenvalues before and after removal of the core electron,
Eq. \ref{ts3} can be verified, showing that to a good
approximation the core eigenvalue indeed drops linearily relative to the 
Fermi energy as charge is removed.

\begin{figure}
\epsfxsize=7cm \centerline{\epsffile{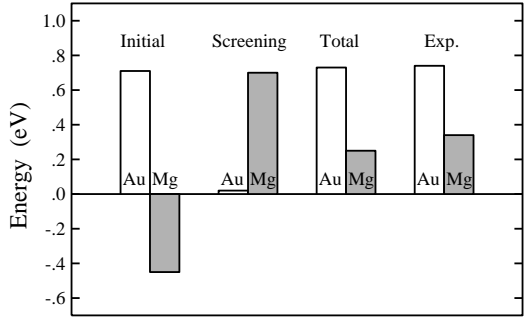}}
\caption{Graphical representation of the calculated and measured core-level 
   binding energy shifts in Table \protect\ref{t2}.
   The full calculated result in column ``Total'' is decomposed
   into the intial-state and screening contributions. 
   Note that the screening contribution on the Mg atom
   drastically changes the picture.}
\label{cls}
\end{figure}

The calculated initial-state and full core-level shifts are obtained 
by taking the differences of the corresponding values in Table \ref{t3},
leading to the values shown in Table \ref{t2} and (graphically)
in Fig.\ref{cls}.
The difference between the full and initial-state CLS
then gives the screening contribution. The full results are in 
good agreement with experiment for both cases.  We find that 
the initial-state estimate is already accurate for the Au 4$f$ shift, 
but that it is grossly incorrect and even has the wrong sign for Mg 1$s$. 
The screening contribution to the shift
is thus completely different for the two types of atom: 
it is negligible for Au, but is the dominant contribution 
for Mg. The screening energies are in reasonable agreement with 
those deduced from Auger parameter measurements, but are incompatible
with the assumptions made in earlier work.\cite{wert}

In view of the discussion in Section \ref{back1}, the conclusion is that
the Au $4f$ core hole is screened equally well in the pure metal
and in the alloy. For Mg, on the other hand, the depopulated 
$1s$ core level has dropped by a larger amount in the alloy,
showing that the core hole is screened significantly less effectively
there than in pure Mg. Given the size of the effect, an analysis of 
measured core level shifts which does not take screening into account 
is pointless.

The calculations  reproduce the experimental core-level 
shifts and split these unambiguously into
an initial-state and a final-state screening term. 
In the rest of this section, we discuss these contributions 
separately in view of the calculated electronic structure 
and the potential-type models.
In the end, we will come to the conclusion that the potential
models  are difficult to justify on the basis of
realistic calculations.

\begin{figure}
\epsfxsize=6cm \centerline{\epsffile{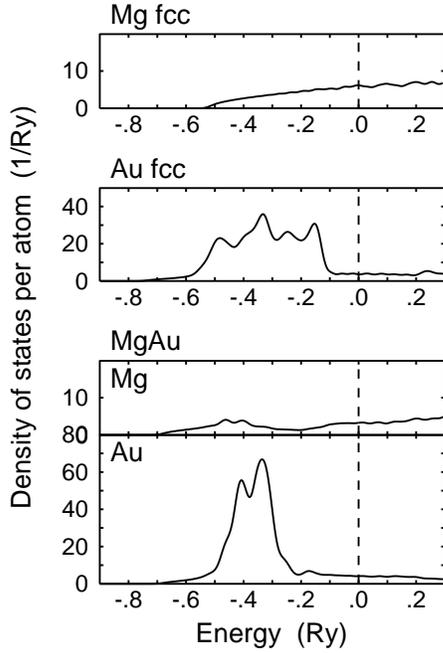}}
\caption{Calculated site-resolved density of states for
   Mg, Au, and MgAu.}
\label{dos1}
\end{figure}

\begin{figure}
\epsfxsize=6cm \centerline{\epsffile{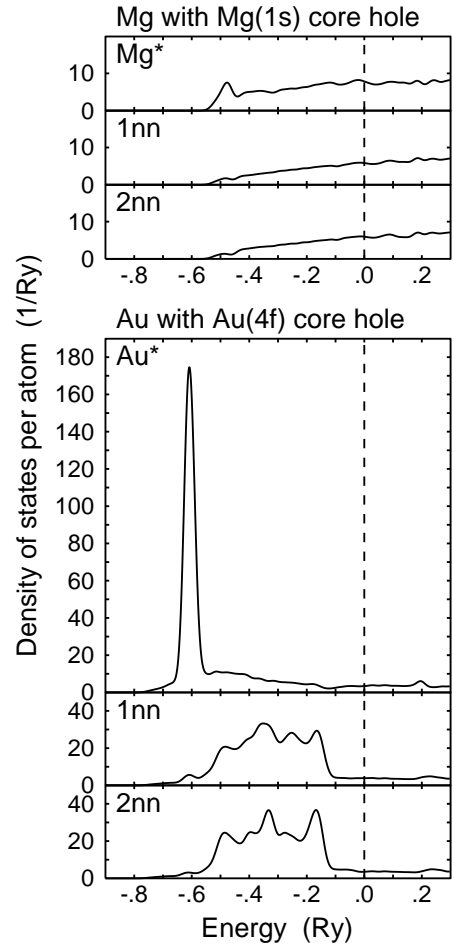}}
\caption{Calculated site-resolved density of states for
   Mg and Au with, respectively,
   a Mg $1s$ or Au $4f$ core hole. Mg$^*$ or Au$^*$ denotes
   the site with the core hole, and 1nn and 2nn are the first
   and second-nearest-neighbor sites, respectively. }
\label{dos2}
\end{figure}

\begin{figure}
\epsfxsize=6cm \centerline{\epsffile{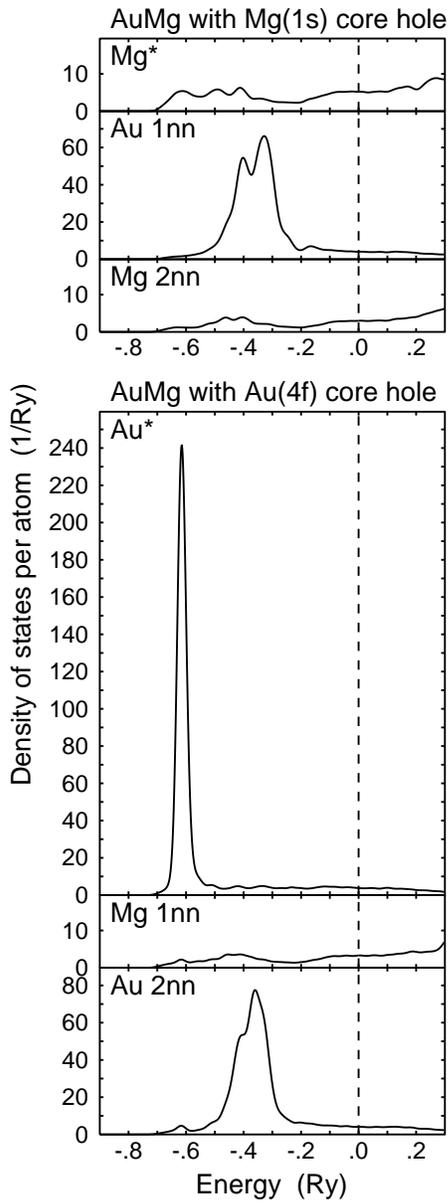}}
\caption{Calculated site-resolved density of states for
 MgAu with 
   a Mg $1s$ or Au $4f$ core hole. Details as in Fig.\protect\ref{dos2}.}
\label{dos3}
\end{figure}

To help in the interpretation of the results, the site-resolved densities 
of states are presented in Figs.~\ref{dos1}, \ref{dos2}, and \ref{dos3}.
Fig.~\ref{dos1} compares the electronic structure
of Mg, Au, and MgAu before making a core hole. 
The effects of a Mg $1s$ or Au $4f$ core hole on the valence states
of Mg and Au are shown in 
Fig.~\ref{dos2}, those in MgAu in
Fig.~\ref{dos3}.

\subsection{Initial-state shifts} 
The calculated initial-state core-level binding-energy 
(i-CLBE) shift upon alloying Mg and Au to form MgAu is 
$\Delta_{\rm Mg}=-0.45$ eV for the Mg $1s$ state and
$\Delta_{\rm Au}=0.71$ eV for the Au $4f$ state.
A charge transfer of about 0.1--0.2 electrons
from Mg to Au is generally considered reasonable, in view of the
electronegativity values of 1.31 and 2.54 for Mg and Au,
respectively. A reliable definition of the charge transfer 
(say, as charge density integral) from a 
density-functional calculation is very difficult to set up, so we 
will not try to verify the generally accepted value directly.
However, we can inspect whether the calculated initial-state shift
is compatible with the potential model for this accepted value of the
charge transfer. We also discuss other modeling concepts which
attempt to explain the initial-state CLBE shift. In the end,
an honest appraisal is that
no simple model can account for the calculated values, despite
extensive efforts to find one.

As described above, the basic feature of the potential model is
that a charge transfer to the atoms of type $B$ causes a
repulsive on-site Coulomb potential which pushes up the core states,
reducing the (initial-state) CLBE. This effect is only partly
compensated by the Madelung potential. Thus, the CLBE is reduced
for those atoms which acquire additional charge, and vice versa.
On the other hand, we can also present an equally simple alternative model,
based on a rigid-band description, where this effect is reversed.
Assume (with reference to Figs.~\ref{dos1}, \ref{dos2}, and \ref{dos3})
 that the density of states
(DOS) of the alloy is obtained by adding together the two DOS 
of the constituents, shifted vertically to line up 
in some way which reflects the bonding.
The charge on an atom in the alloy is then simply related to
the position of the shared alloy Fermi energy with respect to
the site-decomposed valence DOS. 
Furthermore, we assume that the core eigenvalues are at a fixed position
relative to the valence band, so that the core levels track the shifts
of the DOS. Since the CLBE is defined relative to the Fermi energy,
it follows that charge transfer to sites of type $B$ is associated with 
an upward shift of the Fermi energy and a {\it larger} initial-state CLBE.

In the case of MgAu, we are assuming that there is charge transfer from 
Mg to Au, and have calculated that there is an increase of the Au $4f$ CLBE
in the alloy. The Mg atom has lost some charge by alloying
and has a reduced CLBE. Even if only the signs are considered,
these features are incompatible with the potential model, but agree with 
the rigid-band description. In fact, we can become ambitious and
try use the calculated values of the density of states at the Fermi level 
for the pure Mg and Au materials
($D_F$(Mg)=0.46 states/eV and $D_F$(Au)=0.33 states/eV)
to connect the CLBE shifts with the charge transfer.
Assuming a resonably flat DOS at the Fermi energy,
the transferred charge for an atom of a certain type is
approximately equal to
$$
   \Delta q = D_{\rm F}\, \Delta \epsilon_c 
$$ 
which yields $-0.21$ and $0.23$ electrons for Mg and Au, respectively.
These values are close to the generally accepted
charge transfer for this type of alloy; furthermore, 
the charge transferred away from Mg is close to the
charge transferred to the Au atom.

Unfortunately, this gratifying result must be considered
accidental, for several reasons. Foremost is that
the alloy has a substantially smaller volume than
the sum of the volumes of the consituents: the cell volumes are
113.3, 146.9, and 225.9 bohr$^3$ for Au, Mg, and MgAu, 
respectively. It makes sense to
assign the shrinkage of 13\% to the softer Mg atom. Thus, a more
correct description could be to ``prepare'' the Mg atom by
compressing it to a smaller volume, then forming the alloy
from this compressed Mg$'$ and the Au crystal.
The total i-CLBE shift then is a sum of the effects due to
the two steps. 
Independent of whether the shrinkage is assigned to the Mg
or Au atoms, we are now considering a simpler system
in which an alloy is formed without any volume change.
If the rigid-band model is a resonable description, it should be 
equally applicable here.
When the Mg bulk is compressed, the $1s$ core level moves 
up by 0.51~eV, reducing the CLBE from
1248.62~eV to $\epsilon_c$=1248.11~eV. This value is almost equal to
that in the alloy, so that the estimated charge transfer for
the Mg atom from the rigid-band model now comes out close to zero. 
Unfortunately, this is not
compatible with the charge of 0.23 electrons added to the
Au site, throwing the perceived success of the rigid-band model into doubt.

At this stage, it can be speculated that shifts of the Fermi
energy should be included, arising from changes in the electronic
structure due to alloying. Indeed, the plots of the DOS in Figs.~\ref{dos1} 
to \ref{dos3} show that the rigid-band assumption 
is not conspicuously well satisfied.
To demonstrate that all kinds of other effects of similar 
magnitude would still be neglected, we focus on just one aspect,
namely the role of $sp$ to $d$ charge promotion during alloying.
This can be most easily investigated in an ASA calculation,
where the total crystal volume is assigned to atomic spheres.
The ASA result for the initial-state CLBE shifts
when alloying Mg$'$ and Au reproduces the full-potential
calculation reasonably well. The interpretation of the
CLBE shift can now be given a new dimension, since we can directly
investigate the response of the core eigenvalues to changes
in the $sp$ and $d$ charges. We obtain response parameters
$\partial\epsilon_c / \partial Q_\ell \approx 3$ eV/electron
for the $sp$ and $\approx$1.5 eV/electron for the $d$ states.
Furthermore, we can inspect the changes in the partial charges
$Q_\ell$ when the Mg atom is taken from the pure (compressed)
Mg$'$ crystal and is placed in the alloy, obtaining
$\Delta Q_{sp} \approx -0.21$ and $\Delta Q_{d} \approx 0.28$ 
electrons.
Thus, only 0.07 electrons are added to the Mg atomic sphere,
but $\approx 0.25$ electrons are promoted from the $sp$ to the
$d$ states. Combined with the response parameters, it follows
that a contribution to the initial-state CLBE shift of about $-0.3$ eV 
should be attributed to the $sp$ to $d$ promotion.
Altogether, this shows that not only do the core states shift relative
to the valence band when charge is transferred, but the effect is
significantly different depending on the angular momentum which 
takes up the charge. 

In sum, despite attempts in various directions,
we have not been able to find a simple model which can 
describe the initial-state CLBE shifts caused by alloying.
The simple potential model is not applicable because even the signs
are not predicted correctly. The rigid-band model seems
slightly more plausible, but also suffers from
a number of shortcomings. Theoretically, an extended
model could be written down which includes numerous other relevant
effects, such as $sp$-to-$d$ promotion etc. However, this model would
be so complicated and unwieldy that the overall aim of a simple 
model would be negated. Having confidence in our calculations,
we believe that the values of $-0.45$ eV and $0.71$ eV for the
i-CLBE shift are reliable, but we have no convincing 
way to explain these numbers in simple terms.

\subsection{Final-state screening contribution} 
Next, we discuss the final-state screening contribution to the CLBE shifts
when Mg and Au are alloyed to make ordered MgAu.
As mentioned above, the calculated screening contribution is 0.02 eV for Au 
and 0.70 eV for Mg (Table \ref{t1}). This was interpreted as 
follows: the screening of the Au $4f$ core hole happens in 
a way which is nearly independent of the environment.
In contrast, the screening of the Mg $1s$ core hole is significantly
different in the pure Mg bulk and in the alloy. More exactly,
the core hole in the alloy is screened considerably less
effectively than in pure Mg. In the following, we try to 
analyze this difference in the screening properties.

\begin{figure}
\epsfxsize=7cm \centerline{\epsffile{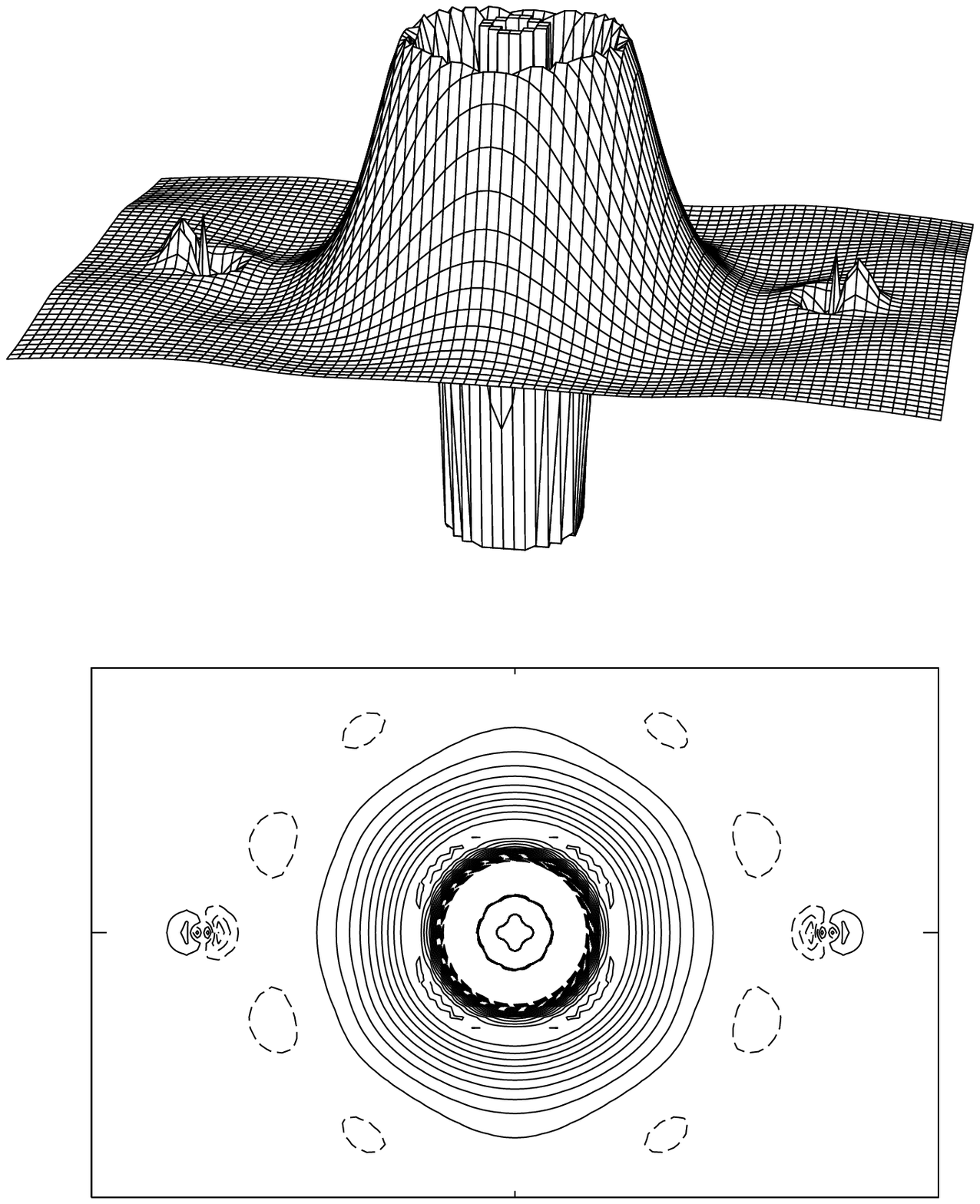}}
\caption{Screening charge density for a 1$s$ core hole at the Mg site in fcc Mg. 
   Contour spacing $\pm 5$, $\pm 15$,\ldots times $10^{-3}$ bohr$^{-3}$.
   The solid and dashed lines show positive and negative densities, 
   respectively.}
\label{f1}
\end{figure}

\begin{figure}
\epsfxsize=7cm \centerline{\epsffile{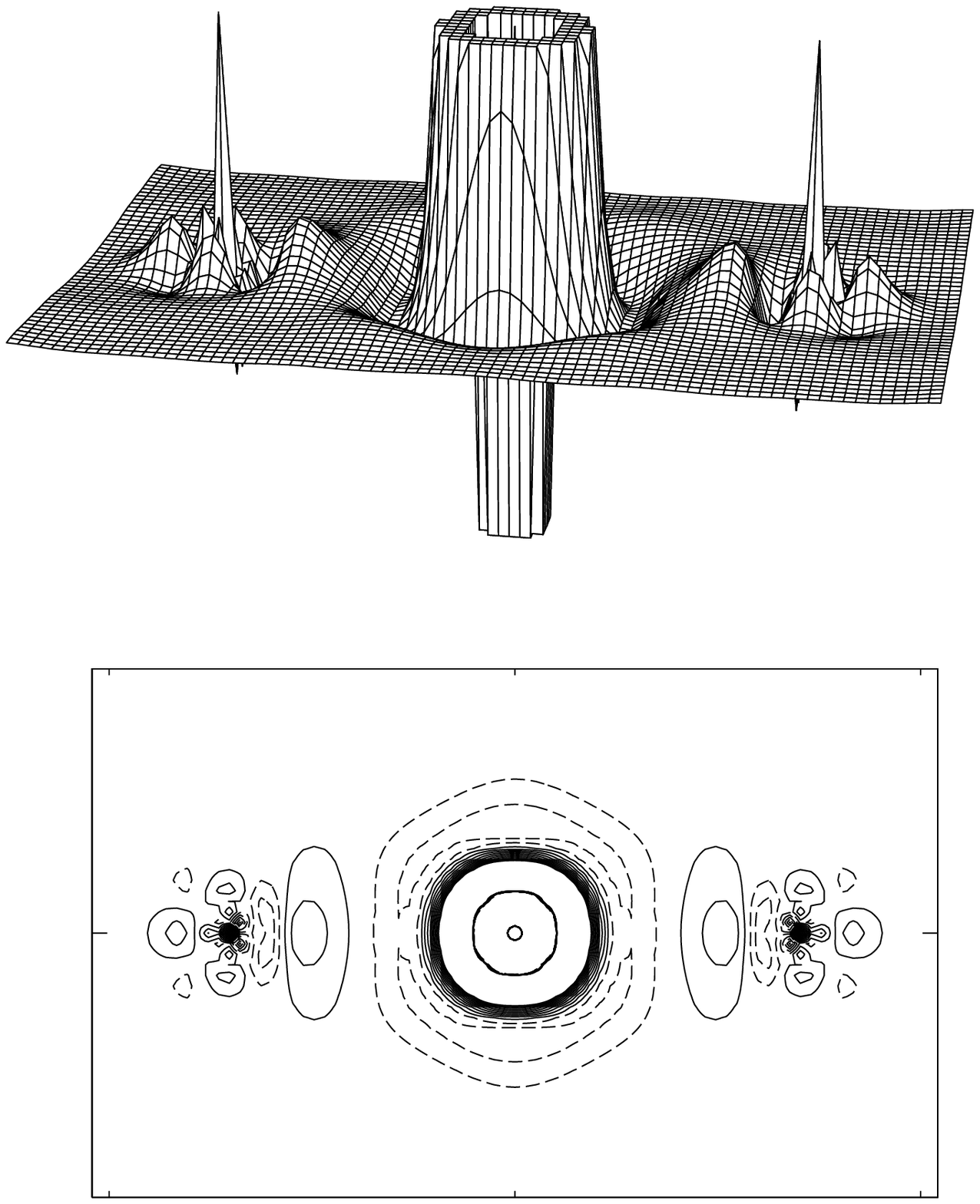}}
\caption{Screening charge density for a 4$f$ core hole at the Au site 
   in fcc Au. Details as in Fig.\protect\ref{f1}.}
\label{f2}
\end{figure}

\begin{figure}
\epsfxsize=7cm \centerline{\epsffile{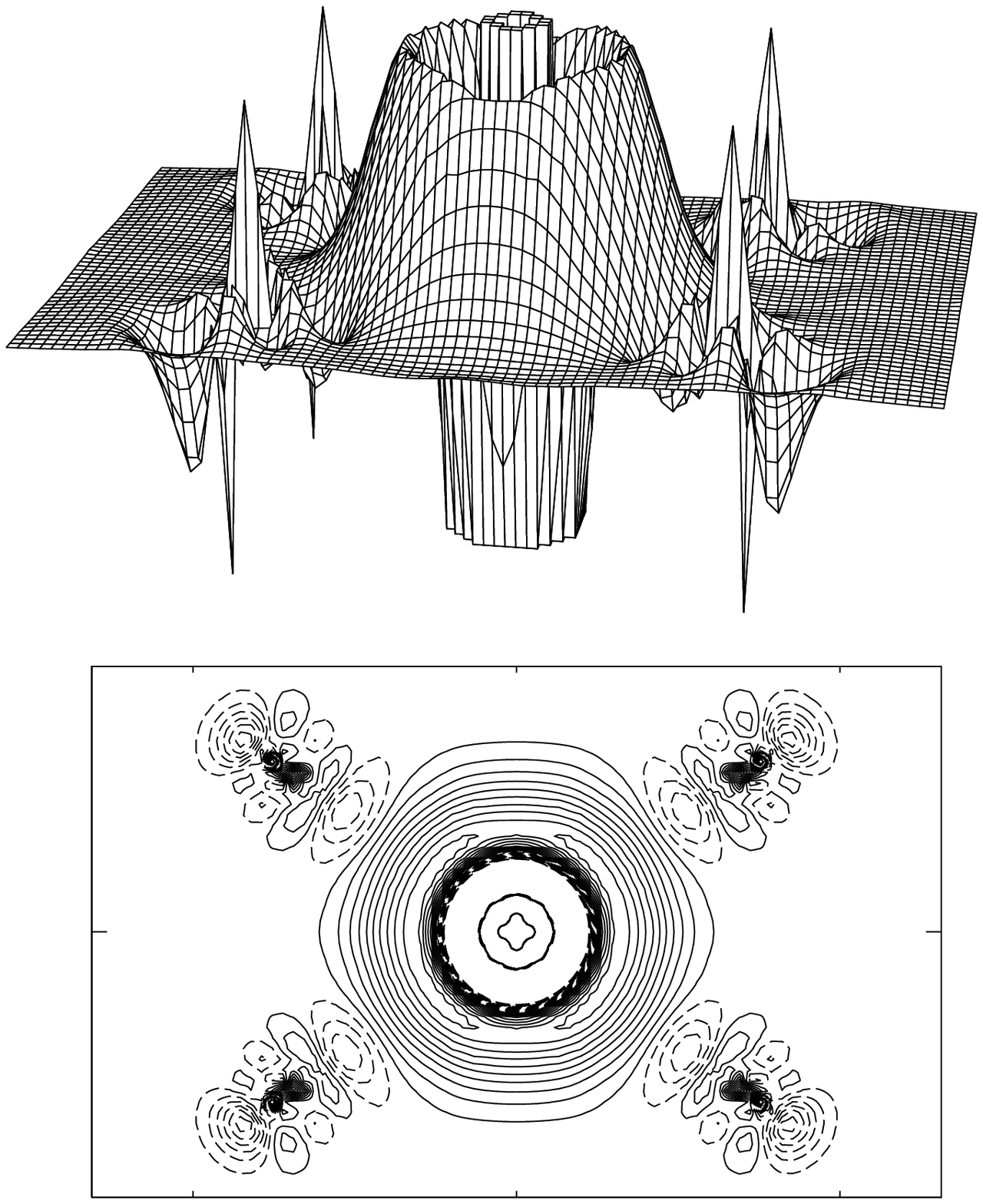}}
\caption{Screening charge density for a 1$s$ core hole at the Mg site 
   in MgAu. Details as in Fig.\protect\ref{f1}.}
\label{f3}
\end{figure}

\begin{figure}
\epsfxsize=7cm \centerline{\epsffile{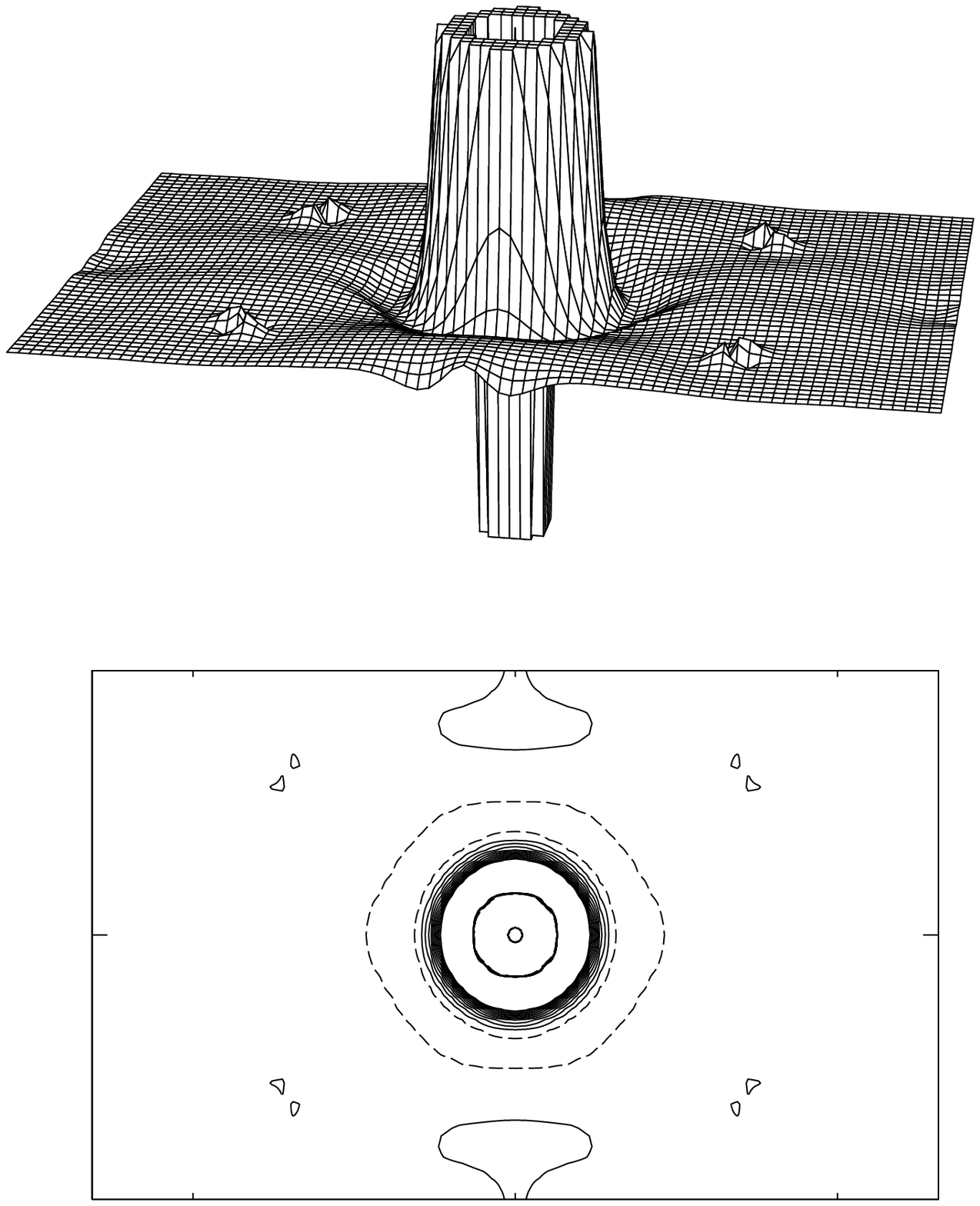}}
\label{f4}
\caption{Screening charge density for a 4$f$ core hole at the Au site 
   in MgAu. Details as in Fig.\protect\ref{f1}.}
\end{figure}

A major advantage of an accurate simulation such as a DFT
calculation is that it can provide data which is not accessible
to experiment; one example is the separation into initial-state
and screening contributions. It is equally useful to use the calculation 
as a ``microscope'' to provide quantities such as the DOS
or the charge density. In the present situation, we can
develop a feeling for the nature of the core-hole screening
by inspecting the screening density directly.
This is simply the difference of the charge density with and
without the core hole. One additional electron is in the
valence charge, responding to the attractive potential of the
core hole more or less flexibly. The screening densities are shown 
in Figs.~\ref{f1}, \ref{f2}, \ref{f3}, and \ref{f4} for pure Mg, pure Au, 
the Mg core hole in MgAu, and the Au core hole in MgAu,
respectively. These plots are a central result of this paper,
making it possible to think about the screening cloud 
in a straightforward and unambiguous way.

By comparing Figs.\ref{f1} and \ref{f2} for the pure constituents,
basic differences for Mg and Au are evident.
Whereas the screening cloud in Mg is wide and extended, 
screening in Au is performed by a localized lump of electrons.
For a true transition metal, this could be easily explained:
the screening electron would be taken up by the localized $d$ states 
at the Fermi energy. For Au, however, the $d$ shell is already full 
and this explanation is not possible. 

Instead, the correct explanation for the localised screening in Au
can be deduced from the corresponding DOS plot.
On the Au$^*$ atom with the core hole, the $d$ states are 
pulled down (and out of the crystal $d$ band) by the attractive
core-hole potential. The electronic structure is similar to that 
of a Hg impurity in Au. In real space, the $d$ states contract,
albeit without any change in the occupation number. To the screening
cloud, this process contributes the difference of the contracted and
uncontracted $d$ shell, which is a positive peak near the nucleus
surrounded by a negative ``ring.'' At this stage, we have not yet
taken up the extra screening electron. This is done by the $sp$ states
which now in turn screen (and fill in) the attractive ring.
Since the $sp$ states are more extended, this cannot be done
completely, leaving some part of the negative ring visible in the
total screening cloud. 

For the quality of the screening, the charge closest to the nucleus
is most relevant. On the Au atom this is dominated by the shrinking
of the $d$ shell, which can be expected to be largely
independent of the environment. Screening by $sp$ electrons
is a more extended affair which can be influenced by the
environment of the atom. However, for Au the $sp$ electrons
play a less immediate role, even though they actually take up
the additional screening electron. In contrast, screening of
the Mg $1s$ core hole is done only by $sp$ valence electrons 
in the form of an extended cloud. Overall, these arguments can explain 
why screening is largely independent of the environment in Au but not in Mg,
in agreement with the calculated results for the alloying process.

Next, we compare the screening of the core holes in the 
pure materials and the MgAu alloy. For the Au $4f$ core hole,
the screening clouds in the density plots look very similar
in the central $d$-electron lump, with some differences in
the outer regions. Based on the discussion above, we can easily
accept that the screening is similar in Au and MgAu
and that only an insignificant contribution to the Au $4f$ CLBE shift
is obtained. For the case of Mg, we wish to understand why the
screening in the alloy is significantly less effective. 
Unfortunately, in this context it is again difficult
to obtain a clear answer.

A first possible explanation for the less effective screening 
in the alloy is that charge has been transferred
away from the Mg atom, leaving less charge to 
respond to the attractive core hole, leading to reduced screening.
However, if we count the number of electrons inside
a sphere of a fixed radius ($R_0=2.8$ bohr) we find that 
the sphere charge in the alloy is 0.36 electrons {\it above} that
in pure Mg. This is presumably a consequence of the reduced
volume in the alloy. Even though the additional charge is mainly
in the outer regions of the sphere, it would seem to 
invalidate an explanation based on reduced available valence charge.

Secondly, a comparison of the DOS for pure Mg with the Mg site in MgAu
shows that the simple-metal parabolic $sp$ DOS has changed 
to some degree of covalent character in the alloy, with
a minimum of the DOS around $-0.2$ Ry. It can be speculated that
this leads to a somewhat more rigid valence charge density, which cannot
respond as flexibly to the core hole potential. This effect could
play a role, but is not confirmed or invalidated 
by the calculation.

Finally, Fig.\ref{f3} shows antiscreening features on the
neighboring Au atoms. This could be interpreted as a
``variable wavelength Friedel oscillation'' whereby the
Friedel wavelength changes from a value appropriate to Mg
to a shorter one on the Au atoms. The antiscreening could 
push the first node of the screening density inwards
with a corresponding reduction of the screening charge.
This explanation, while potentially applicable, also cannot be
confirmed unambiguously. 

\section{Summary and Conclusions} 

In this paper, we have presented the results of {\it ab initio} 
density-functional theory calculations of the core level shifts 
which arise upon alloying, using the prototypical intermetallic
compound MgAu as an example. 
We were interested in the following questions:
how well the experimental results can be reproduced; 
how the full core level shifts can be separated into
initial-state and final state screening contributions; 
and whether the results can be understood in terms of simple models.

The agreement to experiment turns out to be good. 
The calculated core-level shifts are $0.73$ and $0.25$ eV for 
the Au $4f$ and Mg $1s$ states, respectively, close to the measured values
of $0.74$ and $0.34$ eV. Given the complexity of the
problem, these results are very satisfying. We have also found
that the absolute core-level binding energies, calculated
as the difference of two total energies, agree to within 1\%
with the experimental results.

The calculations give an unambiguous separation of each
core level shift into a static initial-state contribution
and a term due to the final-state screening of the
core hole by the other electrons. Such a separation is central
to all subsequent attempts to understand the results using
simpler concepts. Somewhat unexpectedly, we find that the
screening contribution is not just a small correction,
but changes the picture drastically. Specifically, the 
shift of the Mg $1s$ core state changes sign when screening effects
are included. 

Extensive attempts were made to evaluate the calculated
results in terms of simpler models. For the initial-state
shifts, however, no convincing model could be found
which is able to predict the calculated values. 
Among other considered descriptions, the well-known
``simple potential model'' could not be confirmed. 
The basic difficulty is that a large number of effects influence
the core level binding energy. The situation is considerably
too complicated to be cast into any simple model with
only a few parameters. Possibly, a series of calculations
for several different systems could uncover trends and help
to formulate a better model, but we do not consider it plausible that 
an adequate general model can be found.

For the screening contribution, we have used the calculations
to obtain accurate images of the screening clouds for the different cases.
This information cannot be obtained from experiment 
and is of major help when trying to obtain insight into the nature of 
the screening process. Indeed, straightforward interpretations for 
the screening mechanism at the Mg and Au sites could be deduced. 
Whereas the screening of the
Mg core hole is done by a relatively extended $sp$-electron cloud,
screening in Au takes place in a two-step process. First, the
full Au $d$ shell contracts in response to the attractive
core-hole potential, then the $sp$ valence electrons fill up the depletion
ring around the $d$ shell. This description is in line 
with the result that the Mg screening depends on the environment,
while Au screening does not.

Two conclusions are drawn from the results. First, while it
is possible to obtain insight into the electronic structure 
changes upon alloying and the screening behaviour, simple models
which try to connect the alloying core-level shifts with charge transfer
cannot be confirmed. This is mainly due to the complexity of the
real system which is not compatible with a description involving
only a few quantities. Specifically, this means that charge transfer
is only one of several quantities involved, and in fact one
of the most poorly defined ones.
Secondly, a full {\it ab-initio} calculation
can reproduce measured core-level binding energies and their
shifts to very good accuracy. This shows that simpler models are
not actually needed in order to interpret measured values,
where such measurements are used to investigate systems with
unknown properties. Instead, density-functional calculations should be 
used for this purpose.

\narrowtext
\begin{table}    
\begin{tabular}{l|cc}
\multicolumn{1}{c}{ } &
\multicolumn{1}{c}{ a$_0$ (bohr)} &
\multicolumn{1}{c}{ B$_0$ (Mbar)} \\
\tableline
MgAu th.  & 6.09 & 1.05    \\
MgAu exp.   & 6.15  & ---   \\
\tableline
Au th.  & 7.68 & 1.85  \\
Au exp. & 7.70 & 1.73  \\
\tableline
Mg th.  & 8.38 & 0.40    \\
Mg exp.   & 8.46   & ---   \\
\end{tabular}
\vspace{0.7cm}
\caption{Equilibrium lattice constant and bulk modulus for
MgAu (CsCl structure), Au (fcc) and Mg (fcc).
The experimental Mg lattice constant corresponds to the measured
volume per atom in the hcp structure.}
\label{t1}
\end{table}

\narrowtext
\begin{table}    
\begin{tabular}{lrrrr}
Case& $\epsilon_c(0)$ & $\epsilon_c(1)$ & calc. & exp. \\
\hline
Au 4$f$ in Au   &  78.64 & 93.25 & 85.78 & 85.88  \\
Au 4$f$ in MgAu &  79.35 &  94.15 & 86.51 & 86.62  \\
Mg 1$s$ in Mg   &  1248.62 & 1364.74 & 1306.78 & 1303.20  \\
Mg 1$s$ in MgAu &  1248.17 & 1365.35 & 1307.03 & 1303.54  \\
\end{tabular}
\vspace{0.7cm}
\caption{Calculated and measured core-level binding energies in eV,
referred to the Fermi energy. Column headings: 
``$\epsilon_c(0)$'' and ``$\epsilon_c(1)$''
are the calculated eigenvalues before respectively after making the
core hole; ``calc'' denotes the calculated CLBE as a total-energy
difference; ``exp'' gives the 
experimental data from Ref.\protect\onlinecite{darrah}.}

\label{t3}
\end{table}

\narrowtext
\begin{table}  
\begin{tabular}{lrrrr}
Case & initial & screening & full & exp. \\
\hline
Au 4$f$   & 0.71    & 0.02 & 0.73 & 0.74 \\
Mg 1$s$   & $-0.45$ & 0.70 & 0.25 & 0.34 \\
\end{tabular}
\vspace{0.5cm}
\caption{Mg 1$s$ and Au 4$f$ core-level 
shifts in the MgAu alloy with respect to pure Au and Mg in eV.
Column ``initial'' is the initial-state shift,
``screening'' is the final-state screening contribution, 
``full'' is the full calculation, ``exp'' is the
experimental data from Ref.\protect\onlinecite{darrah}.}
\label{t2}
\end{table}


\narrowtext

\end{multicols}
\end{document}